\documentclass[final,5p,times,twocolumn]{elsarticle}

\usepackage{iftex}
\ifPDFTeX
  \usepackage{CJKutf8}
\else
  \usepackage[UTF8,scheme=plain,fontset=fandol]{ctex}
\fi
\usepackage{amsmath,amssymb,bm,mathtools}
\usepackage{graphicx}
\usepackage[colorlinks=true,linkcolor=blue,citecolor=blue,urlcolor=blue]{hyperref}
\graphicspath{{./}}
\biboptions{sort&compress}

\journal{Physics Letters B}

\begin{document}
\ifPDFTeX
  \begin{CJK*}{UTF8}{gbsn}
\fi

\begin{frontmatter}

\title{Renormalization Group Analysis of Pairing Instabilities in Nuclear Fermi Liquids}

\author[ibs]
{Yang Xiao
  \texorpdfstring{(肖杨)}{(Xiao Yang)}}

\author[riken,ibs]{Yixin Guo
  \texorpdfstring{(郭一昕)}{(Guo Yixin)}
  \corref{cor1}}
  \ead{yixin.guo@riken.jp}

\author[ibs]{Youngman Kim\corref{cor1}}
\ead{ykim@ibs.re.kr}

\cortext[cor1]{Corresponding author}

\affiliation[ibs]{
organization={Center for Exotic Nuclear Studies, Institute for Basic Science},
city={Daejeon},
postcode={34126},
country={Republic of Korea}
}

\affiliation[riken]{
organization={RIKEN Nishina Center for Accelerator-Based Science},
city={Wako},
postcode={351-0198},
country={Japan}
}

\begin{abstract}
A nuclear Fermi liquid exhibits competing pairing instabilities in different spin, isospin, and orbital channels. In a Fermi-surface renormalization group (RG) treatment, the channel that develops a pole first is determined not only by its tree-level attraction but also by its one-loop RG coefficient. We illustrate this mechanism in a minimal \(S/P\)-wave model. A spherical Fermi surface establishes the reference competition between the lowest even- and odd-parity interactions. Axial deformation changes the relevant Fermi-surface integrals and lifts the degeneracy between longitudinal and transverse \(P\)-wave components. In isospin-asymmetric matter, neutron--proton Fermi-momentum splitting restricts the simultaneous low-energy contribution of the two species and can terminate the \(np\) running at finite threshold scales. Our calculations are intended as controlled one-loop RG illustrations rather than as quantitative nuclear-matter  calculations. We show how the Fermi-surface geometry and composition can change the ordering of competing pairing instabilities.
\end{abstract}


\end{frontmatter}

\section{Introduction}
\label{sec:intro}

Pairing correlation is a fundamental many-body phenomenon in the finite Fermi system and it governs a wide range of structural and dynamical properties. 
Nuclear pairing in finite nuclei has been extensively studied, and neutron–proton pairing has attracted considerable attention in recent years~\cite{Frauendorf:2014mja, Uzawa:2024sdu, Wang:2024lef}.

 An important aspect of pairing correlations is the competition among different pairing channels and the emergence of competing orders and clustering phenomena. In particular, the competition and coexistence between $S$- and $P$-wave channels constitute a recurring theme in strongly correlated quantum systems.
In condensed-matter physics, such an interplay has been proposed to give rise to exotic superconducting phases such as anapole superconductivity, with possible relevance to UTe$_2$~\cite{Kanasugi2022Comm.Phys.5.1.,Jiao2020Nature579.523--527}.
Similar phenomena also arise in dense nuclear matter, where the $^1S_0$ and $^3P_2$ neutron superfluids, together with their possible coexistence, have been extensively discussed in the context of neutron stars and magnetars~\cite{Takatsuka1993PTP.112.27,Yasui2020PhysRevC.101.055806}.
Beyond these pairing phenomena, the interplay between $S$- and $P$-wave interactions also plays an important role in neutron-rich halo nuclei~\cite{Hammer2017}, may be relevant to tetraneutron systems~\cite{marques2021quest,duer2022observation}, and has been explored in ultracold atomic gases, where it gives rise to rich few- and many-body physics~\cite{Guo2023Phys.Rev.B107.024511,Guo2026Phys.Rev.B113.054512}.

In a Fermi-surface RG description, the interaction between quasiparticles near the Fermi surface is represented by a four-fermion vertex.
An attractive interaction that is marginal at tree level may become marginally relevant by developing a pole in the one-loop renormalization-group flow as the cutoff is lowered.~\cite{Benfatto:1990zz, Polchinski:1992ed, Shankar:1993pf, Casalbuoni:2018haw}. The pole marks the breakdown of the perturbative RG flow in that channel and provides a simple criterion for studying the competition among various pairing channels.

In this Letter, we investigate the competition among pairing channels by determining which channel first develops a pole under the renormalization-group flow.
We employ a minimal S- and P-wave truncation, which provides the simplest framework in which even- and odd-parity pairing channels can compete while keeping the RG structure transparent. Our aim is not to calculate quantitative nuclear pairing gaps, but rather to isolate how the Fermi-surface geometry and the internal composition of the system influence the one-loop RG coefficients and the resulting pole scales.

We first consider a spherical Fermi surface as the reference case for \(S/P\)-wave competition. We then introduce an axially deformed Fermi surface~\cite{Bulgac:1996ntk, Drut:2009ce,Gebremariam:2010ni}, for which the surface measure and the local Fermi velocity modify the angular weighting of the one-loop correction. Finally, we consider a two-fermion system (e.g., neutrons and protons) with Fermi surfaces of different radii. In that case, the \(np\) running depends on whether neutron and proton states can simultaneously lie in the particle--particle or hole--hole sectors within the running energy shell.

These cases are different realizations of the same mechanism. The pole scale of each interaction component depends on its initial strength and on the coefficient in its one-loop RG equation. This coefficient is fixed by a weighted integral over the Fermi surface and, in a multicomponent system, by the simultaneous availability of low-energy intermediate states from the different species. The present approach therefore provides a compact diagnostic of how Fermi-surface geometry and composition can reorder competing pairing instabilities in nuclear systems.
\section{Fermi-surface RG in the BCS channel}
\label{sec:rg}

We denote a quasiparticle energy--momentum variable by \(P=(E,\mathbf p)\), where \(E\) is measured relative to the chemical potential and \(\mathbf p\) is the spatial momentum. Since the Fermi surface is defined in \(\mathbf p\)-space, only the spatial momentum is decomposed as
\begin{equation}
        \mathbf p=\mathbf k_F(\Omega)+\ell_\perp \hat{\mathbf n}(\Omega),
        \qquad
        \xi_{\mathbf p} \equiv \epsilon(\mathbf p)-\mu
        \simeq v_F(\Omega)\ell_\perp .
        \label{eq:decomp}
\end{equation}
  Here, $\Omega$ parametrizes the tangential directions along the Fermi surface, so that the tangential momentum dependence is encoded in $\mathbf k_F(\Omega)$. The vector $\hat{\mathbf n}(\Omega)$ is the outward unit normal to the Fermi surface, while $\ell_\perp$ measures the displacement in the normal direction. 

A Wilsonian RG adopted in \cite{Polchinski:1992ed} integrates out modes in the thin energy shell
\begin{equation}
        \Lambda_E-\mathrm d\Lambda_E<|\xi_{\mathbf p}|<\Lambda_E,
        \label{eq:shell}
\end{equation}
where \(\Lambda_E\) is the running cutoff and  \(\mathrm d\Lambda_E\) is an infinitesimal shell thickness. Under low-energy rescaling, \(E\) and \(\ell_\perp\), or equivalently \(\xi_{\mathbf p}\), scale toward zero, whereas \(\Omega\) remains a label of the Fermi-surface patch.

We work in the zero-total-spatial-momentum BCS channel and consider the scattering process
\begin{equation}
        (\mathbf p,-\mathbf p)\rightarrow(\mathbf p',-\mathbf p') .
\end{equation}
The running interaction is denoted by \(V_{\Lambda_E}(\mathbf p,\mathbf p')\); spin and isospin labels are suppressed until they are needed below. For a fixed Fermi-surface direction \(\Omega_k\), {\bf we} define a shell
\begin{equation}
        \mathcal S_k(\Lambda_E)
        \equiv
        \left\{\ell_\perp:\
        \Lambda_E-\mathrm d\Lambda_E
        <|v_F(\Omega_k)\ell_\perp|<\Lambda_E\right\}.
        \label{eq:normalshell}
\end{equation}

At tree level, the zero-total-spatial-momentum BCS kinematics makes the four-fermion interaction marginal in the Fermi-surface power counting~\cite{Polchinski:1992ed}. To see this explicitly, we consider
\begin{align}
S&\sim\int\prod_{i=1}^4
\,dt\,d\ell_{\perp i}\,d\Omega_i\,
V\,\psi^\dagger\psi^\dagger\psi\psi\,
\,
\delta^{3}(\mathbf p_1+\mathbf p_2-\mathbf p_3-\mathbf p_4).
\end{align}
Under $E,\ell_\perp\to s(E,\ell_\perp)$, the measures, fields, and
energy-conservation delta function together give an overall factor $s$, before accounting for the spatial momentum-conservation delta function. For generic four-fermion scattering, the spatial delta function primarily constrains the unscaled Fermi-surface coordinates $\Omega_i$ and therefore scales as $s^0$.  In the BCS channel, however, the leading Fermi momenta cancel pairwise,
\(\mathbf k_{F,2}=-\mathbf k_{F,1}\) and \(\mathbf k_{F,4}=-\mathbf k_{F,3}\), so that one remaining momentum-conservation condition constrains the radial displacements from the Fermi surface.  Since \(\ell_\perp\to s\ell_\perp\) under the RG rescaling, this constraint contributes \(\delta(s\ell_\perp)=s^{-1}\delta(\ell_\perp)\), which cancels the otherwise positive scaling power of the four-fermion term.  The BCS interaction therefore has net tree-level scaling \(s^0\)~\cite{Polchinski:1992ed}.  This conclusion applies to both the \(S\)- and \(P\)-wave components retained below: their leading momentum dependence is evaluated at finite Fermi momentum and depends only on the unscaled  Fermi-surface coordinates.  In particular, for a spherical Fermi surface,
\(\mathbf p\cdot\mathbf p' =k_F^2\hat{\mathbf p}\cdot\hat{\mathbf p}'+\mathcal O(\ell_\perp)\), so the leading \(P\)-wave structure is also marginal. 

At one loop, the correction from the thin energy shell is
\begin{align}
\delta V_{\Lambda_E}^{(1)}(\mathbf p,\mathbf p')
={}&-i
\int \frac{\mathrm dE'}{2\pi}
\int_{\rm FS}\frac{\mathrm d^2k}{(2\pi)^2}
\int_{\mathcal S_k(\Lambda_E)}\frac{\mathrm d\ell_\perp}{2\pi}
\nonumber\\
&\times V_{\Lambda_E}(\mathbf p,\mathbf k)
V_{\Lambda_E}(\mathbf k,\mathbf p')
\nonumber\\
&\times
\frac{1}{(1+i0^+)(E+E')-v_F(\Omega_k)\ell_\perp}
\nonumber\\
&\times
\frac{1}{(1+i0^+)(E-E')-v_F(\Omega_k)\ell_\perp}.
\label{eq:oneloop_general}
\end{align}
where \(E'\) is the internal energy and
\begin{equation}
        \mathbf k=\mathbf k_F(\Omega_k)+\ell_\perp\hat{\mathbf n}(\Omega_k)
\end{equation}
is the internal spatial momentum.

For a separable partial-wave component \(A\), we write
\begin{equation}
        \mathcal V_A(\mathbf p,\mathbf p';\Lambda_E)
        =g_A(\Lambda_E)V_A(\mathbf p)V_A(\mathbf p'),
        \label{eq:separable}
\end{equation}
where \(g_A\) is the running coupling and \(V_A(\mathbf p)\) denotes the momentum structure of that component. The internal momentum is off the Fermi surface before the shell integration. Expanding its vertex structure, we obtain the following
\begin{equation}
        V_A\!\left(\mathbf k_F+\ell_\perp\hat{\mathbf n}\right)
        =V_A(\mathbf k_F)+\mathcal O(\ell_\perp).
        \label{eq:vertexexpansion}
\end{equation}
After the \(E'\) integration, the shell integral contains \(\int_{\rm shell}\mathrm d\ell_\perp/|\ell_\perp|\), which produces the factor \(\mathrm d\Lambda_E/\Lambda_E\). The first term in Eq.~\eqref{eq:vertexexpansion} contributes to this scale-dependent term, whereas terms containing positive powers of \(\ell_\perp\) are nonlogarithmic or power suppressed. At this order, the internal vertex structure can therefore be evaluated on the Fermi surface.

Substituting Eq.~\eqref{eq:separable} into Eq.~\eqref{eq:oneloop_general}, we have the following
\begin{align}
\delta \mathcal V_A(\mathbf p,\mathbf p')
={}&-g_A^2(\Lambda_E)V_A(\mathbf p)V_A(\mathbf p')
\nonumber\\
&\times\left[
\int_{\rm FS}
\frac{\mathrm d^2k}{(2\pi)^3v_F(\Omega_k)}
V_A^2\!\left(\mathbf k_F(\Omega_k)\right)
\right]
\frac{\mathrm d\Lambda_E}{\Lambda_E}.
\label{eq:deltaVA}
\end{align}
The two factors of \(V_A(\mathbf k_F)\) arise from the two interaction vertices. Matching the correction back to the same momentum structure defines
\begin{equation}
        N_A
        \equiv
        \int_{\rm FS}
        \frac{\mathrm d^2k}{(2\pi)^3v_F(\Omega_k)}
        V_A^2\!\left(\mathbf k_F(\Omega_k)\right),
        \label{eq:NA}
\end{equation}
which is the one-loop RG coefficient in channel \(A\). It is determined by a weighted Fermi-surface integral and does not by itself measure the interaction strength. The RG equation is
\begin{equation}
        \Lambda_E\frac{\mathrm d g_A(\Lambda_E)}{\mathrm d\Lambda_E}
        =N_Ag_A^2(\Lambda_E).
        \label{eq:flowLambda}
\end{equation}
If \(N_A\) is scale independent, its solution is
\begin{equation}
        g_A(\Lambda_E)
        =
        \frac{g_A(\Lambda_0)}
        {1+N_Ag_A(\Lambda_0)\ln(\Lambda_0/\Lambda_E)} ,
        \label{eq:runningcoupling}
\end{equation}
where \(\Lambda_0\) denotes the initial cutoff~\cite{Polchinski:1992ed}. For an attractive initial coupling \(g_A(\Lambda_0)<0\), the one-loop solution has a formal pole at
\begin{equation}
        \Lambda_A^\ast
        =
        \Lambda_0
        \exp\left[-\frac{1}{N_A|g_A(\Lambda_0)|}\right].
        \label{eq:instabilityscale}
\end{equation}
When several channels are present, the channel with the largest \(\Lambda_A^\ast\) develops the first pole as the cutoff is lowered and is identified as the dominant channel,
\begin{equation}
        A_{\rm dominant}=\operatorname*{arg\,max}_{A}\Lambda_A^\ast.
        \label{eq:selection}
\end{equation}

\section{Spherical benchmark: $S$- and $P$-wave competition}
\label{sec:spherical}

We first fix the normalization in an one-component isotropic system. For a spherical Fermi surface,
\begin{equation}
        \mathrm d^2k=k_F^2\,\mathrm d\Omega_{\hat k},
        \qquad
        v_F(\Omega_k)=v_F.
\end{equation}
For the lowest even-parity interaction,
\begin{equation}
        V_S(\mathbf p,\mathbf p';\Lambda_E)=g_S(\Lambda_E),
\end{equation}
Eq.~\eqref{eq:NA} gives
\begin{equation}
        N_S
        =
        \int_{\rm FS}\frac{\mathrm d^2k}{(2\pi)^3v_F}
        =\frac{k_F^2}{2\pi^2v_F}.
        \label{eq:NSiso}
\end{equation}

For the lowest odd-parity interaction, we take
\begin{equation}
        V_P(\mathbf p,\mathbf p';\Lambda_E)
        =g_P(\Lambda_E)\,\mathbf p\cdot\mathbf p'.
        \label{eq:Pvertex_gP}
\end{equation}
When retaining the term proportional to \(\mathrm d\Lambda_E/\Lambda_E\), only the internal momentum is evaluated at its Fermi-surface value; the external momenta remain as the arguments of the operator. The two \(P\)-wave vertices generate
\begin{equation}
        (\mathbf p\cdot\mathbf k)(\mathbf k\cdot\mathbf p')
        \longrightarrow
        k_F^2(\mathbf p\cdot\hat{\mathbf k})
        (\hat{\mathbf k}\cdot\mathbf p').
\end{equation}
Using
\begin{equation}
        \int\frac{\mathrm d\Omega_{ k}}{4\pi}\,
        \hat k_i\hat k_j=\frac{\delta_{ij}}{3},
\end{equation}
one obtains
\begin{equation}
        (\mathbf p\cdot\mathbf k)(\mathbf k\cdot\mathbf p')
        \longrightarrow
        \frac{k_F^2}{3}\,\mathbf p\cdot\mathbf p'.
        \label{eq:Pprojection}
\end{equation}
Matching the correction back to the same operator gives
\begin{equation}
        \Lambda_E\frac{\mathrm d g_P(\Lambda_E)}{\mathrm d\Lambda_E}
        =\frac{k_F^2}{3}N_Sg_P^2(\Lambda_E).
        \label{eq:gPflow}
\end{equation}
Defining the interaction strength on the Fermi surface by
\begin{equation}
        \lambda_P(\Lambda_E)\equiv k_F^2g_P(\Lambda_E),
\end{equation}
we find
\begin{equation}
        \Lambda_E\frac{\mathrm d\lambda_P(\Lambda_E)}{\mathrm d\Lambda_E}
        =\frac{N_S}{3}\lambda_P^2(\Lambda_E).
        \label{eq:Pflow}
\end{equation}
The factor \(1/3\) is the three-dimensional angular projection associated with the convention in Eq.~\eqref{eq:Pvertex_gP}; it is not an additional dynamical suppression of all physical \(P\)-wave interactions. In this convention, the spherical \(S/P\) boundary is
\begin{equation}
        |\lambda_P|=3|g_S|.
        \label{eq:boundary}
\end{equation}

Figure~\ref{fig:isotropic} summarizes the spherical benchmark by comparing the formal pole scales of the \(S\)- and \(P\)-wave couplings. For each point in the \((|g_{S0}|,r_P)\) plane, we define
\begin{equation}
        r_P
        \equiv
        \frac{|k_{F0}^{\,2}g_{P0}|}{|g_{S0}|},
\end{equation}
and identify the component with the larger pole scale.  Since the equality of the two pole scales reduces to \(r_P=3\) in the present normalization, the channel-selection boundary is independent of the overall magnitude of the initial \(S\)-wave attraction.  The \(S\)-wave component is leading for \(r_P<3\), whereas the \(P\)-wave component is leading for \(r_P>3\). The diagram represents the ordering of the one-loop pole scales and should not be interpreted as a thermodynamic phase diagram.

\begin{figure}[t]
\centering
\includegraphics[width=0.96\linewidth]{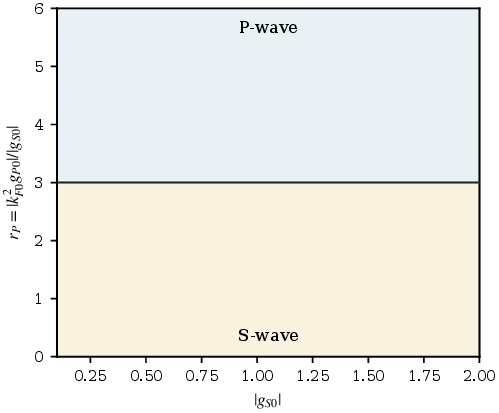}
\caption{
Spherical channel-selection diagram for the minimal \(S/P\)-wave model. 
}
\label{fig:isotropic}
\end{figure}

\section{Axially deformed Fermi surfaces}
\label{sec:axial}

We next keep a single Fermi surface but deform its geometry. 
An axially symmetric surface~\cite{Bulgac:1996ntk, Drut:2009ce,Gebremariam:2010ni} is parameterized as
\begin{equation}
        \mathbf k_F(\theta,\phi)=k_F(\theta)\hat{\mathbf r}(\theta,\phi),
        \qquad
        \xi_{\mathbf p}\simeq v_F(\theta)\ell_\perp.
\end{equation}
Its surface element is
\begin{equation}
\begin{aligned}
        \mathrm d^2k\equiv\mathrm dS
        &=J(\theta)\,\mathrm d\theta\,\mathrm d\phi,\\
        J(\theta)
        &=k_F(\theta)\sin\theta
        \sqrt{k_F^2(\theta)+[k_F'(\theta)]^2}.
\end{aligned}
        \label{eq:Jdef}
\end{equation}
The tree-level scaling of the interaction is unchanged. The deformation enters the one-loop flow through the surface measure, the local Fermi velocity, and the momentum structure of the vertex.

For the \(S\)-wave interaction, the one-loop RG coefficient is
\begin{equation}
        N_S^{\rm ax}
        =\int_{\rm FS}\frac{\mathrm d^2k}{(2\pi)^3v_F(\theta)}
        =\frac{1}{4\pi^2}
        \int_0^\pi\mathrm d\theta\,\frac{J(\theta)}{v_F(\theta)}.
        \label{eq:NSax}
\end{equation}

Axial symmetry splits the lowest odd-parity interaction into longitudinal and transverse components,
\begin{equation}
        V_P(\mathbf p,\mathbf p')
        =g_zp_zp_z'+g_\perp(p_xp_x'+p_yp_y').
        \label{eq:Psplit}
\end{equation}
The corresponding internal momentum components on the Fermi surface are
\begin{equation}
\begin{aligned}
        K_z(\theta,\phi)&=k_F(\theta)\cos\theta,\\
        K_x(\theta,\phi)&=k_F(\theta)\sin\theta\cos\phi,\\
        K_y(\theta,\phi)&=k_F(\theta)\sin\theta\sin\phi.
\end{aligned}
\label{eq:axstructures}
\end{equation}
The mixed term proportional to \(K_xK_y\) vanishes after the azimuthal integration. Moreover, axial symmetry gives identical coefficients for the \(x\) and \(y\) components. We therefore define \(N_\perp\) as the coefficient of either transverse component, which accounts for the factor \(1/2\) below:
\begin{align}
        N_z
        &=\int_{\rm FS}\frac{\mathrm d^2k}{(2\pi)^3v_F(\theta)}K_z^2(\theta,\phi)
        \nonumber\\
        &=\frac{1}{4\pi^2}\int_0^\pi\mathrm d\theta\,
        \frac{J(\theta)k_F^2(\theta)\cos^2\theta}{v_F(\theta)},
        \label{eq:Nz}\\
        N_\perp
        &=\int_{\rm FS}\frac{\mathrm d^2k}{(2\pi)^3v_F(\theta)}
        \frac{K_x^2(\theta,\phi)+K_y^2(\theta,\phi)}{2}
        \nonumber\\
        &=\frac{1}{8\pi^2}\int_0^\pi\mathrm d\theta\,
        \frac{J(\theta)k_F^2(\theta)\sin^2\theta}{v_F(\theta)}.
        \label{eq:Nperp}
\end{align}
In the spherical limit, the three Cartesian \(P\)-wave components are degenerate,
\begin{equation}
        N_z=N_\perp=N_P^{\rm sph}=\frac{k_F^2}{3}N_S.
\end{equation}
Thus \(N_z\) and \(N_\perp\) are not additive pieces of \(N_P^{\rm sph}\); they are the two axial continuations of the same spherical \(P\)-wave eigenvalue.

For a small quadrupole deformation,
\begin{equation}
        k_F(\theta)=k_{F0}[1+\beta_2Y_{20}(\theta)],
        \qquad
        v_F(\theta)=\frac{k_F(\theta)}{m},
        \label{eq:quad}
\end{equation}
one obtains, to first order in \(\beta_2\),
\begin{align}
        N_S^{\rm ax}
        &=N_S^{\rm sph}[1+\mathcal O(\beta_2^2)],
        \label{eq:NSbeta}\\
        N_z
        &=N_P^{\rm sph}\left[1+\frac{3}{5}\sqrt{\frac{5}{\pi}}\,\beta_2
        +\mathcal O(\beta_2^2)\right],
        \label{eq:Nzbeta}\\
        N_\perp
        &=N_P^{\rm sph}\left[1-\frac{3}{10}\sqrt{\frac{5}{\pi}}\,\beta_2
        +\mathcal O(\beta_2^2)\right],
        \label{eq:Nperpbeta}
\end{align}
where
\begin{equation}
        N_S^{\rm sph}=\frac{k_{F0}^2}{2\pi^2v_{F0}},
        \qquad
        N_P^{\rm sph}=\frac{k_{F0}^2}{3}N_S^{\rm sph},
        \qquad
        v_{F0}=\frac{k_{F0}}{m}.
\end{equation}
A positive quadrupole deformation increases the longitudinal coefficient and decreases the transverse coefficient; a negative deformation has the opposite effect.

The corresponding flows are
\begin{equation}
        \Lambda_E\frac{\mathrm d g_i(\Lambda_E)}{\mathrm d\Lambda_E}
        =N_i g_i^2(\Lambda_E),
        \qquad i=S,z,\perp,
        \label{eq:axflows}
\end{equation}
with pole scales
\begin{equation}
        \Lambda_i^\ast
        =\Lambda_0\exp\left[-\frac{1}{N_i|g_i(\Lambda_0)|}\right].
\end{equation}
The dominant channel is the one with the largest \(\Lambda_i^\ast\).

Figure~\ref{fig:axial} shows how axial deformation modifies the competition among the \(S\), \(P_z\), and \(P_\perp\) components.  At each point in the \((\beta_2,r_P)\) plane, the three pole scales are evaluated using the coefficients \(N_S^{\rm ax}\), \(N_z\), and \(N_\perp\), and the component with the largest pole scale is selected.  At \(\beta_2=0\), the two \(P\)-wave components are degenerate and the spherical boundary \(r_P=3\) is recovered.  For positive \(\beta_2\), the longitudinal coefficient \(N_z\) increases while the transverse coefficient \(N_\perp\) decreases, so that the \(P_z\) component is favored relative to \(P_\perp\).  Negative \(\beta_2\) produces the opposite ordering.

The initial longitudinal and transverse \(P\)-wave couplings are taken to be equal. Therefore, the distinction between 
the \(P_z\)- and \(P_\perp\)-dominated regions in Fig.~\ref{fig:axial} arises entirely from the deformation dependence of the one-loop RG coefficients.

The diagram is intended as a schematic comparison of
pole scales; a quantitative treatment would require diagonalizing the full interaction kernel in each fixed-\(m\) sector.

\begin{figure}[t]
\centering
\includegraphics[width=0.96\linewidth]{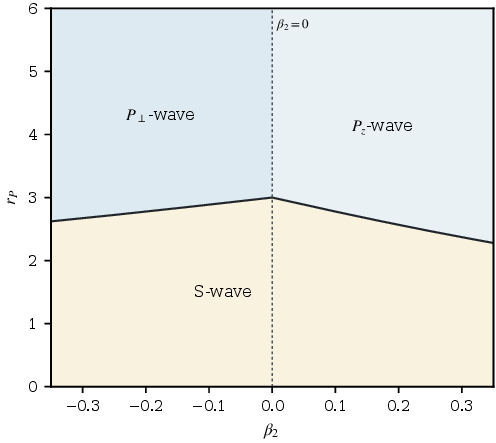}
\caption{
Schematic \(S/P_z/P_\perp\) competition on an axially deformed Fermi surface, obtained from Eqs.~\eqref{eq:NSax}--\eqref{eq:Nperp}. The plot uses \(k_{F0}=m=1\) and equal initial \(P_z\) and \(P_\perp\) strengths.}
\label{fig:axial}
\end{figure}

\section{Fermi-momentum splitting and neutron--proton pairing competition}
\label{sec:isospin}

We now retain a spherical geometry but allow the neutron and proton Fermi momenta to differ. Their linearized dispersions are
\begin{equation}
        \xi_\alpha(k)=v_{F,\alpha}(k-k_{F,\alpha}),
        \qquad \alpha=n,p,
        \label{eq:xispecies}
\end{equation}
where the two internal lines carry spatial momenta \(\mathbf k\) and \(-\mathbf k\), with \(k=|\mathbf k|\). 

For identical-particle channels, the one-loop running retains the one-component form, with species-dependent RG coefficients
\begin{align}
        N_\alpha^{(S)}
        &=\frac{k_{F,\alpha}^2}{2\pi^2v_{F,\alpha}},
        \label{eq:NalphaS}\\
        N_\alpha^{(P)}
        &=\frac{k_{F,\alpha}^2}{3}N_\alpha^{(S)}
        =\frac{k_{F,\alpha}^4}{6\pi^2v_{F,\alpha}},
        \qquad \alpha=n,p.
        \label{eq:NalphaP}
\end{align}
For nonrelativistic quasiparticles, \(v_{F,\alpha}=k_{F,\alpha}/m_\alpha^\ast\); for comparable effective masses, the larger Fermi surface therefore gives the larger identical-particle RG coefficient.

The neutron--proton contribution is more restrictive because the two internal states are measured relative to different Fermi surfaces. Its internal-energy integral is
\begin{align}
        \mathcal I_{np}(E,k)
        ={}&-i\int\frac{\mathrm dE'}{2\pi}
        \frac{1}{(1+i0^+)(E+E')-\xi_n(k)}
        \nonumber\\
        &\times\frac{1}{(1+i0^+)(E-E')-\xi_p(k)}.
        \label{eq:np_energy_integral}
\end{align}
The two poles in the complex \(E'\) plane lie on opposite sides of the real axis when \(\xi_n(k)\) and \(\xi_p(k)\) have the same sign. When the signs are opposite, the contribution vanishes. The contour integral gives
\begin{equation}
        \mathcal I_{np}(E,k)
        =\frac{1-\theta[-\xi_n(k)]-\theta[-\xi_p(k)]}
        {\xi_n(k)+\xi_p(k)-2E}.
        \label{eq:npkernelE}
\end{equation}
For the infinitesimal shell flow, \(|E|\ll |\xi_\alpha(k)|\sim\Lambda_E\), and hence
\begin{equation}
        \mathcal I_{np}(k)
        \simeq\frac{1-\theta[-\xi_n(k)]-\theta[-\xi_p(k)]}
        {\xi_n(k)+\xi_p(k)}.
        \label{eq:npkernel}
\end{equation}
The numerator retains the particle--particle and hole--hole sectors and removes the mixed-sign sector.

For definiteness, let \(k_{F,n}>k_{F,p}\) and define \(\Delta k_F=k_{F,n}-k_{F,p}\). The mixed-sign interval is \(k_{F,p}<k<k_{F,n}\); the allowed hole--hole and particle--particle regions are \(k<k_{F,p}\) and \(k>k_{F,n}\), respectively. At a running cutoff \(\Lambda_E\), the complete low-energy radial window of species \(\alpha\) is
\begin{equation}
        I_\alpha(\Lambda_E)=
        \left[k_{F,\alpha}-\frac{\Lambda_E}{v_{F,\alpha}},
        k_{F,\alpha}+\frac{\Lambda_E}{v_{F,\alpha}}\right].
\end{equation}
The two full windows begin to overlap at
\begin{equation}
        \Lambda_{\rm ov}
        =\frac{v_{F,n}v_{F,p}}{v_{F,n}+v_{F,p}}\,\Delta k_F.
        \label{eq:LambdaOverlap}
\end{equation}
This unrestricted overlap is not sufficient for the \(np\) running, because it may lie entirely in the mixed-sign interval removed by Eq.~\eqref{eq:npkernel}. Simultaneous particle--particle or hole--hole states within the two low-energy windows require the stronger conditions
\begin{equation}
        \Lambda_E>\Lambda_{\rm pp}\equiv v_{F,p}\Delta k_F,
        \qquad
        \Lambda_E>\Lambda_{\rm hh}\equiv v_{F,n}\Delta k_F.
        \label{eq:np_threshold_scales}
\end{equation}
The scales \(\Lambda_{\rm pp}\) and \(\Lambda_{\rm hh}\) are therefore the particle--particle and hole--hole threshold scales. Above \(\Lambda_{\rm pp}\), the particle-like portions of the two windows overlap; above \(\Lambda_{\rm hh}\), their hole-like portions overlap. By contrast, \(\Lambda_{\rm ov}\) marks only the onset of unrestricted shell overlap. For \(k_{F,p}>k_{F,n}\), the neutron and proton labels are interchanged.

For later use, it is convenient to introduce the pair-energy variable
\begin{equation}
        \eta(k)=\xi_n(k)+\xi_p(k)
        =(v_{F,n}+v_{F,p})(k-k_\ast),
        \label{eq:eta}
\end{equation}
where
\begin{equation}
        k_\ast=\frac{v_{F,n}k_{F,n}+v_{F,p}k_{F,p}}
        {v_{F,n}+v_{F,p}}.
        \label{eq:kstar}
\end{equation}
The point \(k_\ast\) is the zero of the pair-energy denominator; for unequal Fermi momenta it lies in the blocked mixed-sign interval and is not itself an allowed intermediate state. Its role is to define the radial variable \(\eta\) and the corresponding Jacobian. Approximating the smooth radial measure by its value at \(k_\ast\) gives the reference RG coefficients
\begin{equation}
        N_{np}^{(S)}
        =\frac{k_\ast^2}{\pi^2(v_{F,n}+v_{F,p})},
        \qquad
        N_{np}^{(P)}=\frac{k_\ast^2}{3}N_{np}^{(S)}.
        \label{eq:Nnp}
\end{equation}
Within this symmetric radial-weight approximation, the particle--particle and hole--hole sectors each contribute one half of \(N_{np}^{(A)}\). In the isospin-symmetric limit, \(k_\ast=k_F\) and \(v_{F,n}=v_{F,p}=v_F\), and Eq.~\eqref{eq:Nnp} reduces to the one-component result.

The scale-dependent \(np\) flow can then be represented as
\begin{align}
        \Lambda_E\frac{\mathrm d g_{np}^{(A)}(\Lambda_E)}{\mathrm d\Lambda_E}
        ={}&\frac{N_{np}^{(A)}}{2}
        \left[\Theta(\Lambda_E-\Lambda_{\rm pp})
        +\Theta(\Lambda_E-\Lambda_{\rm hh})\right]
        \nonumber\\
        &\times\left[g_{np}^{(A)}(\Lambda_E)\right]^2,
        \qquad A=S,P.
        \label{eq:np_flow_thresholds}
\end{align}
The  numerator in Eq.~\eqref{eq:npkernel} removes the mixed-sign sector. The two step functions in Eq.~\eqref{eq:np_flow_thresholds} switch off the particle--particle and hole--hole contributions below their respective threshold scales. Assuming \(\Lambda_0\) lies above both thresholds, integration gives
\begin{align}
        \frac{1}{g_{np}^{(A)}(\Lambda_E)}
        ={}&\frac{1}{g_{np}^{(A)}(\Lambda_0)}
        +\frac{N_{np}^{(A)}}{2}
        \ln\frac{\Lambda_0}{\max(\Lambda_E,\Lambda_{\rm pp})}
        \nonumber\\
        &+\frac{N_{np}^{(A)}}{2}
        \ln\frac{\Lambda_0}{\max(\Lambda_E,\Lambda_{\rm hh})}.
        \label{eq:np_running_thresholds}
\end{align}

For an attractive initial coupling, we define the unrestricted reference pole scale
\begin{equation}
        \Lambda_{np,A}^{\ast,\mathrm{unres}}
        =\Lambda_0\exp\left[-\frac{1}{N_{np}^{(A)}
        |g_{np}^{(A)}(\Lambda_0)|}\right].
        \label{eq:np_unrestricted_pole}
\end{equation}
Let
\begin{equation}
        \Lambda_>\equiv\max(\Lambda_{\rm pp},\Lambda_{\rm hh}),
        \qquad
        \Lambda_<\equiv\min(\Lambda_{\rm pp},\Lambda_{\rm hh}).
\end{equation}
Solving \(1/g_{np}^{(A)}(\Lambda_E)=0\) gives
\begin{equation}
        \Lambda_{np,A}^{\ast}
        =
        \begin{cases}
        \Lambda_{np,A}^{\ast,\mathrm{unres}},
        &
        \Lambda_{np,A}^{\ast,\mathrm{unres}}\geq\Lambda_>,
        \\[1.2ex]
        \displaystyle
        \frac{\left[\Lambda_{np,A}^{\ast,\mathrm{unres}}\right]^2}{\Lambda_>},
        &
        \sqrt{\Lambda_<\Lambda_>}\leq
        \Lambda_{np,A}^{\ast,\mathrm{unres}}<\Lambda_>,
        \\[2ex]
        \text{no pole},
        &
        \Lambda_{np,A}^{\ast,\mathrm{unres}}<
        \sqrt{\Lambda_<\Lambda_>}.
        \end{cases}
        \label{eq:np_actual_pole}
\end{equation}
In the nearly symmetric case, \(v_{F,n}\simeq v_{F,p}\simeq\bar v_F\), the two cutoffs are approximately equal,
\begin{equation}
        \Lambda_{\rm pp}\simeq\Lambda_{\rm hh}
        \equiv\Lambda_{\rm th}
        \simeq\bar v_F|k_{F,n}-k_{F,p}|.
\end{equation}
The intermediate branch in Eq.~\eqref{eq:np_actual_pole} then disappears: an \(np\) pole occurs only when the unrestricted reference pole lies above the common threshold scale.

To illustrate the competition among the different particle and partial-wave channels, we parameterize the neutron and proton Fermi momenta as
\begin{equation}
        k_{F,n}=k_{F0}(1+\delta),
        \qquad
        k_{F,p}=k_{F0}(1-\delta),
        \label{eq:delta_parameterization}
\end{equation}
and take \(m_n^\ast=m_p^\ast=m\).  The identical-particle interactions are chosen to have the same initial strengths,
\begin{equation}
        g_{nn,A}(\Lambda_0)
        =
        g_{pp,A}(\Lambda_0)
        \equiv g_{A0},
        \qquad A=S,P,
\end{equation}
whereas the neutron--proton interactions are parameterized as
\begin{equation}
        g_{np,A}(\Lambda_0)
        =
        \eta_{np}g_{A0}.
\end{equation}
For the illustrative diagram, we use
\begin{equation}
        k_{F0}=m=\Lambda_0=1,
        \qquad
        N_S^{\rm sph}|g_{S0}|=1,
        \qquad
        \eta_{np}=1.5.
        \label{eq:np_figure_parameters}
\end{equation}

{The choice \(\eta_{np}=1.5\) is an illustrative enhancement of the neutron-proton interaction. Its qualitative physical motivation is that an \(np\) pair can access both \(T=1\) and \(T=0\) isospin channels, whereas \(nn\) and \(pp\) pairs are restricted to \(T=1\). In realistic nuclear systems, the \(T=0\) neutron-proton interaction can be more attractive, as exemplified by the deuteron channel \(^{3}S_{1}\!-\!{}^{3}D_{1}\). We therefore take \(\eta_{np}>1\).}

{The particular value \(\eta_{np}=1.5\), however, is not fitted to a realistic nuclear force and should not be interpreted as a quantitative prediction. In the present model it is chosen to generate a visible finite region in which the enhanced \(np\) attraction overcomes the suppression caused by the neutron-proton Fermi-momentum mismatch. Applying the same enhancement to both the schematic \(S\)- and \(P\)-wave couplings is also a simplifying assumption rather than a statement about their microscopic interaction strengths.}

{For \(\eta_{np}=1\), the \(nn\), \(pp\), and \(np\) pole scales are degenerate at \(\delta=0\) within the present model, as expected when their initial couplings and Fermi surfaces are identical. At nonzero \(\delta\), the \(np\) running is additionally restricted by the particle-particle and hole-hole threshold scales, whereas the identical-particle channels are not. }

At every point in the \((\delta,r_P)\) plane, we compare the valid pole scales of the six components
\[
        nn:S,\quad nn:P,\quad
        pp:S,\quad pp:P,\quad
        np:S,\quad np:P.
\]
An \(np\) component is included in this comparison only when Eq.~\eqref{eq:np_actual_pole} yields a pole before its running terminates.

The resulting channel-selection diagram is shown in Fig.~\ref{fig:np}.  Near \(\delta=0\), the neutron and proton Fermi momenta coincide and the particle--particle and hole--hole threshold
scales vanish.  The \(np\) components can therefore continue to run to lower scales and, for the enhanced initial \(np\) attraction used here, they provide the leading channels around the symmetric point.

As \(|\delta|\) increases, the Fermi-momentum splitting raises \(\Lambda_{\rm pp}\) and \(\Lambda_{\rm hh}\), thereby reducing the range over which the \(np\) coupling continues to run.  The leading channel then changes to \(nn\) pairing on the neutron-rich side
(\(\delta>0\)) and to \(pp\) pairing on the proton-rich side (\(\delta<0\)).  Increasing \(r_P\) favors the \(P\)-wave components, whereas smaller \(r_P\) favors the corresponding \(S\)-wave components. The curvature of the \(S/P\) boundaries reflects the dependence of the \(P\)-wave RG coefficients on \(k_{F,n}\), \(k_{F,p}\), and \(k_\ast\). The visible corners occur where three competing pole scales become equal and should not be interpreted as numerical artifacts.

\begin{figure}[t]
\centering
\includegraphics[width=0.96\linewidth]{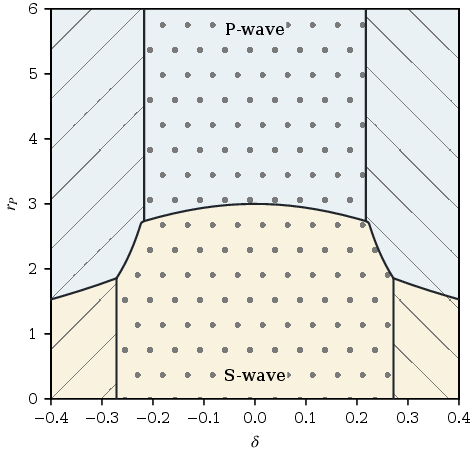}
\caption{Dominant pairing channel in isospin-asymmetric matter as a function of the Fermi-momentum splitting $\delta$.  Pairing composition is distinguished by texture: dots denote $np$ pairing, upward-sloping hatching (left-hand side) denotes $pp$ pairing, and downward-sloping hatching (right-hand side) denotes $nn$ pairing.}
\label{fig:np}
\end{figure}

\section{Conclusions}
\label{sec:conclusions}

The three calculations considered here show how the structure of the Fermi surface changes the one-loop RG coefficients and the associated pole scales. In the spherical benchmark, the relative \(S/P\) running is fixed by the angular projection of the internal momentum. Axial deformation changes the weighted Fermi-surface integrals and lifts the degeneracy between the longitudinal \(P_z\) and transverse \(P_\perp\) components. In isospin-asymmetric matter, the neutron--proton Fermi-momentum splitting restricts the simultaneous low-energy particle--particle and hole--hole contributions and introduces finite threshold scales.

Consequently, Fermi-surface geometry and neutron--proton Fermi-momentum splitting can change which channel develops a pole first even when the initial interaction strengths are held fixed. The comparison is therefore controlled jointly by the initial interaction and by the corresponding one-loop RG coefficient.

The present \(S/P\) basis is intended only as a minimal illustration of the competition between even- and odd-parity interactions. A realistic nuclear calculation would require a multi-partial-wave treatment including uncoupled channels, such as \({}^1S_0\), \({}^3P_0\), and \({}^3P_1\), as well as coupled channels, such as \({}^3S_1-{}^3D_1\) and \({}^3P_2-{}^3F_2\), with the interactions provided by realistic nuclear forces~\cite{Wiringa:1994wb,Machleidt:2000ge,Entem:2003ft,Epelbaum:2004fk,Lu:2021gsb}.

The termination of the zero-momentum \(np\) running below the splitting-induced threshold scales also motivates an extension to pairs with nonzero total momentum. A finite pair momentum may improve the patchwise alignment of the neutron and proton Fermi surfaces, providing an RG realization of a Fulde–Ferrell–Larkin–Ovchinnikov(FFLO)-type instability~\cite{PhysRev.135.A550,LOpaper}. 
Studying the corresponding finite-momentum pairing kernel is a natural continuation of the present work.

\section*{Declaration of competing interest}
The authors declare no known competing financial interests or personal relationships that could have appeared to influence the work reported in this paper.

\section*{Data availability}
No new experimental data were generated. The numerical illustrations are schematic and can be reproduced from the equations and parameters specified in the text.

\section*{Declaration of generative AI and AI-assisted technologies in the manuscript preparation process}
During preparation of this draft, the authors used ChatGPT to assist with the language editing. The authors reviewed and edited the content and take full responsibility for the manuscript.

\section*{Acknowledgements}
Y.X. is supported by the National Natural Science Foundation of China under Grants No.12347113,  No.12505096, the Chinese Postdoctoral Science Foundation under Grants No.2022M720360. 
Y.G. is supported by RIKEN Special Postdoctoral Researchers Program.  
This work was supported in part by the Institute for Basic Science (IBS-R031-D1) and the National Research Foundation of Korea funded by Ministry of Science and ICT (RS-2024-00436392).

\bibliographystyle{elsarticle-num}
\bibliography{ref}

\ifPDFTeX
  \end{CJK*}
\fi
\end{document}